%% file: neurips_2026.tex
\documentclass{article}
\usepackage{xcolor}

\usepackage{float}
\PassOptionsToPackage{numbers, sort&compress}{natbib}

\usepackage[preprint]{neurips_2026}

\usepackage[utf8]{inputenc} 
\usepackage[T1]{fontenc}    
\usepackage{xcolor}
\definecolor{blue1}{HTML}{2E86AB}
\usepackage[colorlinks=true,
            linkcolor=red,
            citecolor=blue1,
            filecolor=blue1,
            urlcolor=blue1]{hyperref}
\usepackage{url}            
\usepackage{booktabs}       
\usepackage{amsfonts}       
\usepackage{nicefrac}       
\usepackage{microtype}      
\usepackage{xcolor}         
\usepackage{multirow}
\usepackage{tabularx}
\usepackage{array}
\usepackage{graphicx}
\usepackage{amsmath}
\usepackage{booktabs}
\usepackage{makecell}
\usepackage{threeparttable}
\usepackage{wrapfig}
\usepackage{float}
\usepackage[ruled,vlined,linesnumbered]{algorithm2e}
\usepackage{xcolor}
\title{AnisoLift: Anisotropic Latent Representations for Coarse Particle Liquid Enhancement}

\author{%
  Zhengqing Gao$^{1,2}$\thanks{Equal contribution. $^{\dagger}$Co-corresponding authors.
  This work was conducted during the Zhengqing Gao's internship at Shanghai AI Lab, under the guidance of Dr. Huaxi Huang.}~,
  Huaxi Huang$^{1*}$,
  Runqi Lin$^{3}$,
  Yuanyuan Wang$^{2}$,
  Meng Li$^{1,4}$,
  Xi Zhou$^{1}$,\\
  \textbf{Tongliang Liu$^{2,6}$},
  \textbf{Mingming Gong$^{2,5\dagger}$},
  \textbf{Xiao Sun$^{1\dagger}$}\\
  $^{1}$Shanghai Artificial Intelligence Laboratory,\\ $^{2}$Mohamed bin Zayed University of Artificial Intelligence,\\ $^{3}$University of Oxford, $^{4}$The Hong Kong University of Science and Technology, \\$^{5}$The University of Melbourne, $^{6}$The University of Sydney
}

\begin{document}

\maketitle


\begin{abstract}
Particle-based liquid simulation is widely used in graphics and physical modeling, but high-resolution rollouts remain computationally expensive. Consequently, many methods aim to recover fine-scale dynamics and dense transport patterns from coarse particle simulations. However, these methods typically rely on additional particle generation, which still incurs considerable computational overhead and leads to poor representation. To this end, we propose \textbf{AnisoLift}, a structured latent closure framework that augments each coarse particle with learnable anisotropic ellipsoidal components. This allows the model to capture directional local structure from the underlying high-resolution flow without introducing extra particles. Given a coarse simulation, our model predicts residual corrections to particle states to bring the updated state closer to the aligned high-resolution teacher. Our training objective jointly supervises particle dynamics and anisotropic geometric structure, encouraging both physical consistency and structural coherence. Extensive experiments show that our approach enhances coarse liquid simulations through improving fidelity to fully resolved flow behavior.
\end{abstract}

\input{sections/introduction_updated}
\input{sections/related}
\input{sections/method-updated}
\input{sections/experiments}
\input{sections/conclusion}

\newpage
\bibliographystyle{unsrtnat} 
\bibliography{references}

\clearpage

\newpage

\end{document}

%% file: sections/introduction_updated.tex
\section{Introduction}
\label{intro_updated}
\begin{wrapfigure}{r}{0.62\textwidth}
    \centering
    \vspace{-8pt}
    \includegraphics[width=0.60\textwidth]{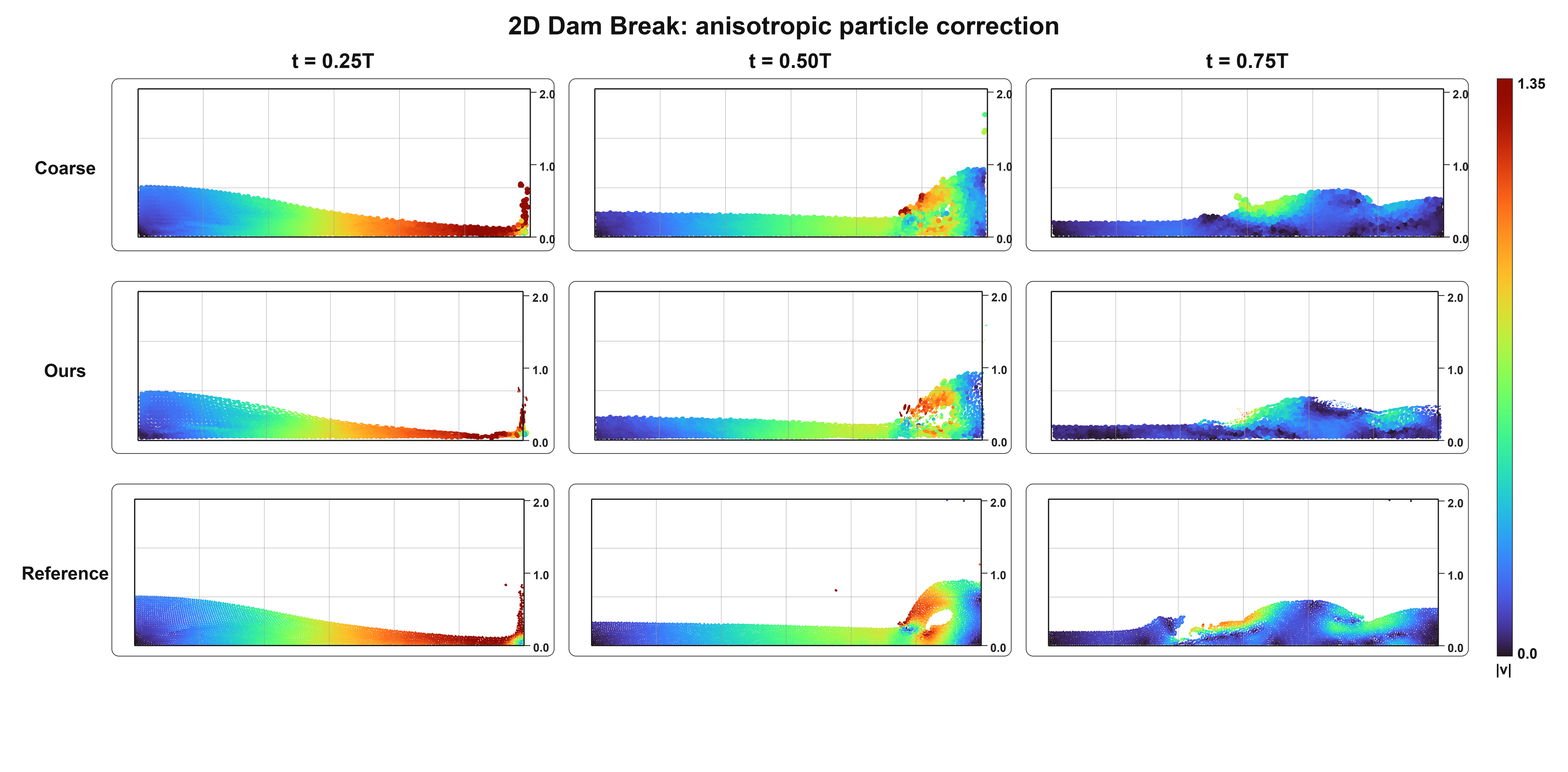}
    \caption{Qualitative correction on the 2D dam-break scene. Coarse particles, our corrected anisotropic coarse state, and the high-resolution reference are shown at two time instants, with particles colored by velocity magnitude.}
    \label{fig:intro_vis}
    \vspace{-8pt}
\end{wrapfigure}
Particle-based liquid simulation is widely used in graphics and physical modeling for the ability to represent free-surface motion and complex liquid interactions \cite{well,fluidgaussian,neuralmpm,extmps,SPH,MPS,particle_fluid_sim,KBST2022,APIC,MLS_MPM,MPM,DPD}. However, the high-quality results are grounded in high-resolution rollouts, which in turn remain computationally expensive, especially over long time horizons and large simulation domains \cite{high_resolution_fluid_sim,CO_FLIP,neuralfluid}. In light of this, coarse particle has been applied in practice. Nevertheless, they usually fail to resolve the fine-scale motion, interface detail, and local structure that distinguish high-quality liquid behavior. As a result, a gap remains between the efficiency of the coarse simulation and the fidelity of fully resolved liquid behavior \cite{um2018mlflip,NeuralUpFlow}.
To reduce this gap, existing work has largely taken two directions. One focuses on learned particle simulators, using graph-based or continuous-kernel models to improve rollout accuracy and long-horizon stability in Lagrangian fluid dynamics \cite{GNS,ummenhofer2020continuousconv,prantl2022dmcf,toshev2024neuralsph}. These methods primarily aim to approximate the underlying particle dynamics, seeking learned update rules accurately over extended rollouts with stability, but they largely overlook the limited expressiveness of a coarse particle representation. The other addresses the problem through coarse-to-high enhancement, seeking to restore fine-scale liquid detail from coarse simulations \cite{NeuralUpFlow,um2018mlflip}. Although these methods target visual and structural recovery, they typically introduce additional particles. Consequently, neither direction succeeds in enhancing coarse liquid simulation quality under a fixed particle budget.


Our starting point is that coarse solvers often preserve the global motion of the flow, while the particles lack the capacity to encode directional local structures present in the high-resolution state. This suggests enriching the local state carried by each coarse particle generated by solvers like Smoothed Particle Hydrodynamics (SPH) \cite{SPH} and Moving Particle Simulation (MPS) \cite{MPS},  rather than increasing the number of particles. We therefore introduce \textbf{AnisoLift}, a coarse particle liquid enhancement framework built around anisotropic latent representations. During training, aligned high-resolution trajectories provide supervision for learning both particle-level residual corrections and anisotropic local structure. Each coarse particle carries ellipsoidal latent components that encode directional geometry beyond an isotropic point sample, while the model predicts residual corrections on top of a stable coarse solver. Joint supervision on residual motion and anisotropic geometry encourages the learned component to capture unresolved local structure while retaining the coarse solver as the transport backbone. As shown in Fig. \ref{fig:intro_vis}, our method enhanced the quality of the coarse significantly.

Our experiments on Lagrangian 2D and 3D datasets demonstrate that AnisoLift improves coarse particle simulations. We show that per-particle enrichment can improve simulation fidelity, particularly when coarse particles under-resolve interface structure and local transport. Our contributions are summarized as follows,
\begin{itemize}
    \item We formulate fixed-budget coarse-particle liquid enhancement, where the goal is to improve coarse simulations without particle densification or differentiable solver rollout.
    \item We propose AnisoLift, a structured latent closure that augments each coarse particle with current-frame residual corrections and an anisotropic ellipsoidal footprint.
    \item We introduce coarse-supported reference alignment, which constructs particle-wise supervision by locally aggregating reference particles around each coarse particle.
    \item We validate AnisoLift on 2D and 3D particle benchmarks generated by SPH and MPS solvers, and analyze its efficiency, ablations, and resolution scaling behavior.
\end{itemize}

%% file: sections/related.tex
\section{Related Work}
\label{sec:related}

\paragraph{Particle-based liquid simulation.}
Particle-based methods are widely used for liquid simulation because they naturally handle free surfaces, large deformation, splashes, and topology changes.
Classical Lagrangian solvers, including SPH \citep{monaghan1992sph,monaghan2005sph, jax-sph} and MPS \citep{MPS}, represent liquids as moving material samples and update particle states through local interactions.
Graphics-oriented particle and particle-grid methods, such as WCSPH \citep{becker2007wcsph}, PCISPH \citep{solenthaler2009pcisph}, APIC \citep{APIC}, FLIP/PIC methods \citep{particle_fluid_sim}, and MPM-based approaches \citep{MPM,MLS_MPM}, further improve stability, incompressibility, and visual fidelity.
However, high-quality results usually require dense particle sampling and long rollouts, making fully resolved simulations expensive.
Coarse simulations are therefore attractive, but their resolved state is often dominated by sparse particle samples and local interaction neighborhoods.
This limits the ability of a coarse particle set to represent unresolved directional structures such as thin sheets, stretched interfaces, and local transport patterns.
In contrast, our work focuses on this representational bottleneck rather than replacing the solver used to generate the coarse trajectory.

\paragraph{Learned particle dynamics and detail enhancement.}
Learning-based methods have been developed to improve particle simulation accuracy and efficiency.
Early work estimates particle accelerations from local features \citep{ladicky2015regressionforests}, while continuous convolution models \citep{ummenhofer2020continuousconv}, graph network simulators \citep{GNS}, and Neural SPH \citep{toshev2024neuralsph} learn particle update rules from data.
These approaches primarily aim to approximate particle dynamics over rollouts, but the particles are usually still represented as point samples connected through isotropic neighborhoods or learned graph interactions.
Another related direction recovers fine-scale liquid detail from coarse simulations.
Methods such as MLFLIP \citep{um2018mlflip} and Neural UpFlow \citep{NeuralUpFlow} improve visual richness or transport detail by generating additional particles or fine-scale structures.
In contrast, AnisoLift studies a fixed particle-count setting: the explicit particle set remains unchanged, while unresolved sub-particle structure is encoded through anisotropic latent components attached to existing particles.

\paragraph{Anisotropic and Gaussian representations.}
Anisotropic representations have been explored in both particle methods and graphics.
ASPH \citep{owen1998asph2,shapiro1996asph} replaces scalar smoothing lengths with tensor-valued kernels to adapt particle support to direction-dependent resolution. Anisotropic kernels are used for liquid surface reconstruction and thin-structure preservation \citep{yu2013anisotropickernel}.
These methods show the value of ellipsoidal support, but the anisotropy is typically derived from local particle statistics, solver heuristics, or reconstruction objectives.
Recent Gaussian representations further demonstrate the expressiveness of anisotropic primitives, including 3D Gaussian Splatting for radiance-field rendering \citep{kerbl2023gaussiansplatting} and Gaussian Fluids for continuous velocity-field simulation \citep{xing2025gaussianfluids}.
Note that anisotropic components AnisoLift is different from both anisotropic kernels and Gaussian field representations. We attach components to coarse Lagrangian particles, learned from aligned high-resolution trajectories, and used as a structured closure for residual state correction.
This allows a coarse particle to carry unresolved directional structure without spawning additional particles or modifying the solver.

%% file: sections/method-updated.tex
\providecommand{\new}[1]{{\color{blue}#1}}

\section{Methodology}
\begin{figure}
    \centering
    \includegraphics[width=0.95\linewidth]{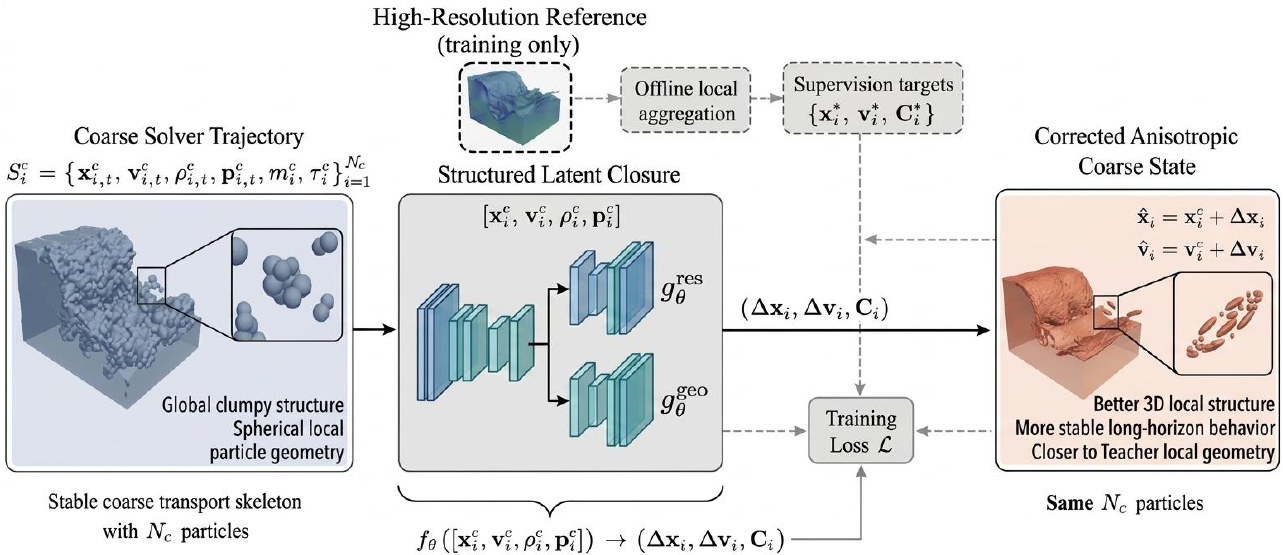}
    \caption{
Overview of AnisoLift. Given a precomputed coarse trajectory, the model predicts current-frame residuals $(\Delta x_i,\Delta v_i)$ and anisotropic geometry $C_i$ from coarse particle features. The corrected state preserves the same coarse particle count while enriching each particle with an anisotropic footprint. High-resolution reference particles are used only during training to construct local supervision targets, with no differentiable solver rollout or particle densification.
}

    \label{fig:placeholder}
\end{figure}
\subsection{Problem Formulation}
\label{sec:problem_formulation}

We consider particle-based liquid simulation in 2D and 3D, with spatial
dimension \(d\in\{2,3\}\). For each scene, we assume paired coarse and
high-resolution reference sequences sampled at the same physical times
\(\{t_0,\ldots,t_{T-1}\}\). The coarse state is
\begin{equation}
\label{eq:coarse_state}
    \mathcal{S}^{c}_{t}
    =
    \left\{
    \left(
    \mathbf{x}^{c}_{i,t},
    \mathbf{v}^{c}_{i,t},
    \rho^{c}_{i,t},
    p^{c}_{i,t},
    m^{c}_{i},
    \tau^{c}_{i}
    \right)
    \right\}_{i=1}^{N_c},
\end{equation}
where \(\mathbf{x}^{c}_{i,t},\mathbf{v}^{c}_{i,t}\in\mathbb{R}^{d}\) denote
particle position and velocity, \(\rho^{c}_{i,t}\) and \(p^{c}_{i,t}\) are
density and pressure, \(m^{c}_{i}\) is mass, and
\(\tau^{c}_{i}\in\{0,1\}\) is a binary type indicator (1 for fluid, 0 for
non-fluid). The reference state is
\begin{equation}
\label{eq:reference_state}
    \mathcal{S}^{r}_{t}
    =
    \left\{
    \left(
    \mathbf{x}^{r}_{j,t},
    \mathbf{v}^{r}_{j,t},
    m^{r}_{j},
    \tau^{r}_{j}
    \right)
    \right\}_{j=1}^{N_r},
    \qquad
    N_c \ll N_r .
\end{equation}

Let \(\mathcal{F}^{c}=\{i:\tau^{c}_{i}=1\}\) and
\(\mathcal{F}^{r}=\{j:\tau^{r}_{j}=1\}\) denote the fluid-particle index sets
in the coarse and reference states. When wall or obstacle particles are
present, they remain in \(\mathcal{S}^{c}_{t}\) and are available to the
model through their attributes and type labels, but supervision is applied only
to fluid particles. The reference state is used only to construct supervision
signals and evaluation targets; our method does not roll out the reference
simulation or require a differentiable solver.

Rather than generating a denser particle set, we keep the coarse particle set
fixed and learn a current-frame closure that corrects the coarse state toward
the high-resolution reference. Given \(\mathcal{S}^{c}_{t}\), the model
predicts residual corrections \((\Delta\mathbf{x}_{i,t},\Delta\mathbf{v}_{i,t})\)
for each fluid particle \(i\in\mathcal{F}^{c}\), yielding
\begin{equation}
\label{eq:corrected_state}
    \widehat{\mathbf{x}}_{i,t}
    =
    \mathbf{x}^{c}_{i,t}+\Delta \mathbf{x}_{i,t},
    \qquad
    \widehat{\mathbf{v}}_{i,t}
    =
    \mathbf{v}^{c}_{i,t}+\Delta \mathbf{v}_{i,t}.
\end{equation}
The model also predicts an anisotropic covariance
\(\mathbf{C}_{i,t}\in\mathbb{S}_{++}^{d}\), which provides each coarse
particle with a local geometric descriptor of unresolved reference structure
(Section~\ref{sec:structured_latent_closure}). Unlike next-step prediction,
these residuals are applied only to the current observed frame and are not fed
back into the solver. The coarse trajectory is generated beforehand by a
stable particle solver, while the learned closure improves coarse-state
expressiveness at the same temporal resolution as the reference data.

\subsection{Paired Trajectory Generation}
\label{sec:paired_trajectory_generation}

For each benchmark scene, we construct paired low- and high-resolution
simulation sequences using the same underlying particle solver. The coarse
sequence and the high-resolution reference sequence are generated independently
at different spatial resolutions, but are sampled at identical physical times.
Specifically, both sequences share the same simulation duration $t_{\mathrm{end}}$,
output interval, and number of saved frames $T$:
\begin{equation}
\label{eq:time_grid}
    t_k = k \Delta t_{\mathrm{out}},
    \qquad
    k = 0,\ldots,T-1,
\end{equation}
where
\begin{equation}
\label{eq:time_grid_def}
    \Delta t_{\mathrm{out}}
    =
    \Delta t_{\mathrm{solver}} \cdot n_{\mathrm{write}},
    \qquad
    T
    =
    \frac{t_{\mathrm{end}}}{\Delta t_{\mathrm{out}}} + 1 .
\end{equation}
Here $\Delta t_{\mathrm{solver}}$ is the solver integration step and
$n_{\mathrm{write}}$ is the number of solver steps between two saved frames.
This construction ensures that $\mathcal{S}^{c}_{t_k}$ and
$\mathcal{S}^{r}_{t_k}$ are temporally aligned for every frame $t_k$.

The coarse sequence
$\{\mathcal{S}^{c}_{t_k}\}_{k=0}^{T-1}$ is produced by a stable solver rollout
at a reduced spatial resolution. It serves as the transport skeleton on which
our model operates. The corresponding high-resolution reference sequence
$\{\mathcal{S}^{r}_{t_k}\}_{k=0}^{T-1}$ provides supervision for learning
current-frame corrections and local geometric structure. Since the two
simulations use different spatial resolutions, their particle counts generally
satisfy $N_c \ll N_r$ and their particles do not have one-to-one
correspondence.

After generating paired sequences, the solver is no longer invoked
during model training. Each training sample is a saved frame
$\left(\mathcal{S}^{c}_{t_k}, \mathcal{S}^{r}_{t_k}\right)$ from the dataset,
rather than a state produced by recursively applying learned corrections
through a differentiable simulator. This decoupled setting has two practical
advantages. First, the learned residuals cannot destabilize the coarse solver
rollout, because they are not fed back into the simulation process. Second, the
learning problem focuses on improving the representation of each coarse frame
under the same temporal resolution as the reference data.

\subsection{Coarse-Supported Reference Alignment}
\label{sec:coarse_supported_reference_alignment}

\begin{figure}
    \centering
    \includegraphics[width=\linewidth]{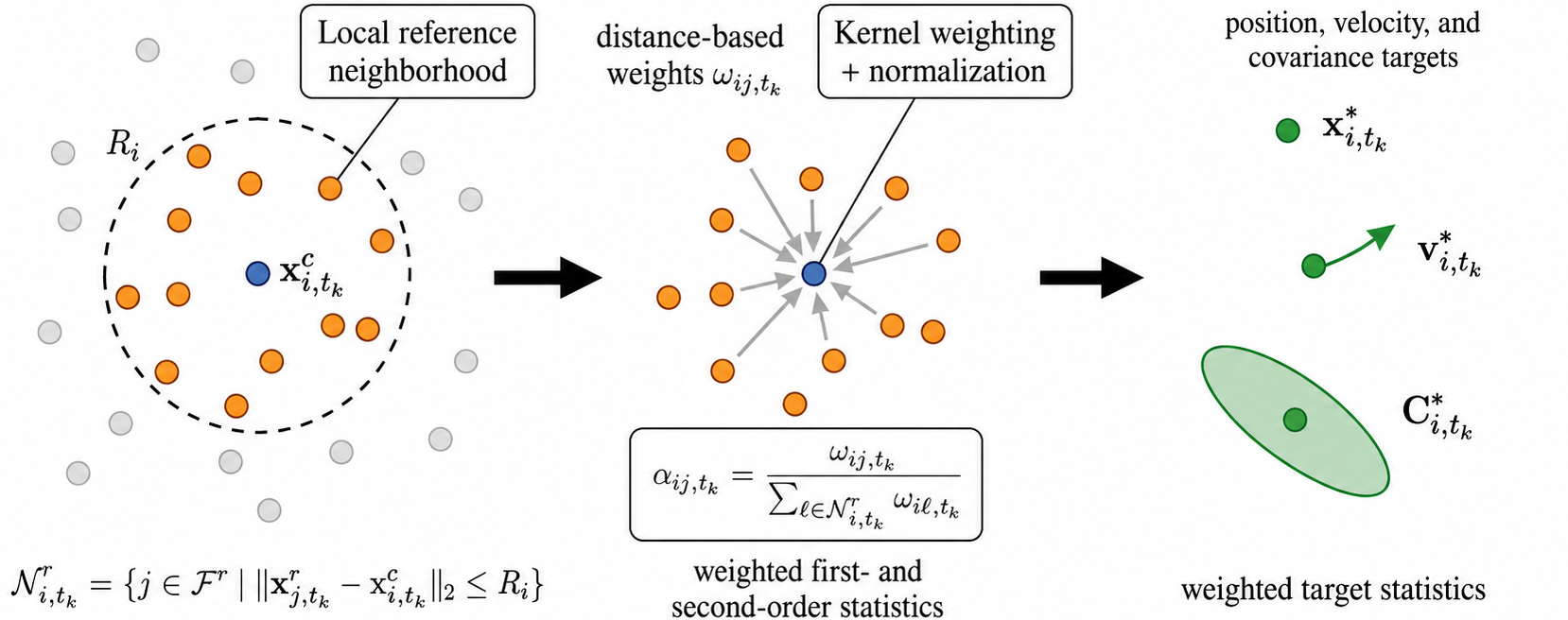}
    \caption{
Coarse-supported reference alignment. Since coarse and high-resolution particles avoid one-to-one correspondence, we construct supervision targets on the coarse support. For each coarse particle \(\mathbf{x}_{i,t_k}^c\), nearby reference particles within radius \(R_i\) are selected, weighted by a distance-based kernel, and aggregated into aligned position, velocity, and covariance targets \((\mathbf{x}_{i,t_k}^*, \mathbf{v}_{i,t_k}^*, \mathbf{C}_{i,t_k}^*)\).
}
    \label{fig:flow_chart}
\end{figure}

Since the coarse and reference simulations use different spatial resolutions,
their particles do not admit one-to-one correspondence. Directly comparing
$\mathbf{x}^{c}_{i,t}$ with an arbitrary reference particle
$\mathbf{x}^{r}_{j,t}$ is therefore ill-defined. We instead construct
coarse-supported reference targets by locally aggregating the high-resolution
reference state around each coarse particle, as shown in Fig \ref{fig:flow_chart}.

For each saved frame $t_k$ and coarse fluid particle $i\in \mathcal{F}^{c}$, we define a local
reference neighborhood:
\begin{equation}
\label{eq:reference_neighborhood}
    \mathcal{N}_{i,t_k}^{r}
    =
    \left\{
    j\in \mathcal{F}^{r}
    \;\middle|\;
    \left\|
    \mathbf{x}^{r}_{j,t_k}
    -
    \mathbf{x}^{c}_{i,t_k}
    \right\|_2
    \leq
    R_i
    \right\},
\end{equation}
where $R_i$ is the support radius associated with the coarse particle. In
practice, $R_i$ is chosen from the coarse particle spacing and solver smoothing
scale.

We assign each reference neighbor a distance-based kernel weight:
\begin{equation}
\label{eq:kernel_weight}
    \omega_{ij,t_k}
    =
    K
    \left(
    \frac{
    \left\|
    \mathbf{x}^{r}_{j,t_k}
    -
    \mathbf{x}^{c}_{i,t_k}
    \right\|_2
    }{R_i}
    \right),
    \qquad
    j \in \mathcal{N}_{i,t_k}^{r},
\end{equation}
where $K(\cdot)$ is a compactly supported radial kernel. The normalized
weights are:
\begin{equation}
\label{eq:normalized_weight}
    \alpha_{ij,t_k}
    =
    \frac{
    \omega_{ij,t_k}
    }{
    \sum_{\ell \in \mathcal{N}_{i,t_k}^{r}}
    \omega_{i\ell,t_k}
    }.
\end{equation}

In practice, a small constant $\epsilon$ is added to the denominator for numerical robustness against vanishing weight sums.
The aligned reference position and velocity for coarse particle $i$ are then
computed as:
\begin{equation}
\label{eq:aligned_targets}
    \mathbf{x}^{*}_{i,t_k}
    =
    \sum_{j \in \mathcal{N}_{i,t_k}^{r}}
    \alpha_{ij,t_k}
    \mathbf{x}^{r}_{j,t_k},
    \qquad
    \mathbf{v}^{*}_{i,t_k}
    =
    \sum_{j \in \mathcal{N}_{i,t_k}^{r}}
    \alpha_{ij,t_k}
    \mathbf{v}^{r}_{j,t_k}.
\end{equation}
These targets provide frame-wise supervision for the corrected coarse state
$\widehat{\mathbf{x}}_{i,t_k}$ and $\widehat{\mathbf{v}}_{i,t_k}$.

We extract a local second-order structure from the same reference neighborhood. The mean target captures the local reference center, while the covariance target preserves the spatial extent and anisotropy that would be lost by a pointwise target alone. The covariance target is:
\begin{equation}
\label{eq:covariance_target}
    \mathbf{C}^{*}_{i,t_k}
    =
    \sum_{j \in \mathcal{N}_{i,t_k}^{r}}
    \alpha_{ij,t_k}
    \left(
    \mathbf{x}^{r}_{j,t_k}
    -
    \mathbf{x}^{*}_{i,t_k}
    \right)
    \left(
    \mathbf{x}^{r}_{j,t_k}
    -
    \mathbf{x}^{*}_{i,t_k}
    \right)^{\top}
    +
    \epsilon_{\mathrm{geo}}\mathbf{I}.
\end{equation}
Here $\epsilon_{\mathrm{geo}}>0$ is a small constant. The diagonal jitter $\epsilon_{\mathrm{geo}}\mathbf{I}$ ensures that
$\mathbf{C}^{*}_{i,t_k}$ is positive definite. This covariance represents the
anisotropic spatial extent of the high-resolution reference region associated
with coarse particle $i$. The result of this step is a set of coarse-indexed supervision signals:
\begin{equation}
\label{eq:frame_targets}
    \mathcal{Y}^{*}_{t_k}
    =
    \left\{
    \left(
    \mathbf{x}^{*}_{i,t_k},
    \mathbf{v}^{*}_{i,t_k},
    \mathbf{C}^{*}_{i,t_k}
    \right)
    \right\}_{i \in \mathcal{F}^{c}},
\end{equation}
which is indexed by the coarse fluid set $\mathcal{F}^{c}$. This makes it possible to train a fixed-size
coarse-particle model even when the reference simulation contains many more
particles.

\subsection{Structured Latent Closure with Anisotropic Ellipsoids}
\label{sec:structured_latent_closure}

Given the coarse state $\mathcal{S}^{c}_{t_k}$ at a frame $t_k$, our
model predicts a structured closure for each coarse particle. The closure has
two coupled components: a current-frame state residual and an anisotropic
ellipsoidal latent representation. The residual corrects the coarse particle
center, while the ellipsoid represents the local spatial extent of the reference region associated with that coarse particle.

For each coarse fluid particle $i\in\mathcal{F}^{c}$, we first construct a feature vector:
\begin{equation}
\label{eq:encoder}
    \mathbf{h}_{i,t_k}
    =
    \phi_{\theta}
    \left(
    \mathcal{S}^{c}_{t_k}, i
    \right),
\end{equation}
where $\phi_{\theta}$ denotes the learnable feature extractor operating on the
current coarse state. The input features are derived from the coarse particle
attributes defined in Section~\ref{sec:problem_formulation}, including
position, velocity, density, pressure, mass, and particle type.

The residual head predicts whitened current-frame corrections:
\begin{equation}
\label{eq:residual_head}
    \left(
    \bar{\Delta\mathbf{x}}_{i,t_k},
    \bar{\Delta\mathbf{v}}_{i,t_k}
    \right)
    =
    g_{\theta}^{\mathrm{res}}
    \left(
    \mathbf{h}_{i,t_k}
    \right),
\end{equation}
where the per-component normalization statistics $(\boldsymbol{\mu}_{\Delta x},\boldsymbol{\sigma}_{\Delta x})$ and $(\boldsymbol{\mu}_{\Delta v},\boldsymbol{\sigma}_{\Delta v})$ are estimated once from the training-split residual targets $\Delta\mathbf{x}^{*}_{i,t_k}=\mathbf{x}^{*}_{i,t_k}-\mathbf{x}^{c}_{i,t_k}$ and $\Delta\mathbf{v}^{*}_{i,t_k}=\mathbf{v}^{*}_{i,t_k}-\mathbf{v}^{c}_{i,t_k}$, induced by the coarse-supported reference targets from Section~\ref{sec:coarse_supported_reference_alignment}. The physical residuals are recovered by
\begin{equation}
\label{eq:denormalize}
\Delta\mathbf{x}_{i,t_k}
    =
    \bar{\Delta\mathbf{x}}_{i,t_k}\odot\boldsymbol{\sigma}_{\Delta x}+\boldsymbol{\mu}_{\Delta x},
    \qquad
    \Delta\mathbf{v}_{i,t_k}
    =
    \bar{\Delta\mathbf{v}}_{i,t_k}\odot\boldsymbol{\sigma}_{\Delta v}+\boldsymbol{\mu}_{\Delta v}.
\end{equation}
where $\odot$ denotes the Hadamard product.
The corrected particle state is then defined by Eq.~\eqref{eq:corrected_state}.
This is a current-frame closure: the predicted residual is used to improve the
state at $t_k$ and is not integrated forward by the particle solver.

To enrich the representation of each coarse particle, we also predict an
anisotropic covariance matrix:
\begin{equation}
\label{eq:geometry_head}
    \mathbf{C}_{i,t_k}
    =
    g_{\theta}^{\mathrm{geo}}
    \left(
    \mathbf{h}_{i,t_k}
    \right),
    \qquad
    \mathbf{C}_{i,t_k} \in \mathbb{S}_{++}^{d},
\end{equation}
where $\mathbb{S}_{++}^{d}$ denotes the space of $d\times d$ symmetric
positive definite matrices. In practice, positive definiteness is enforced by
predicting a matrix factor and forming:
\begin{equation}
\label{eq:covariance_prediction}
    \mathbf{C}_{i,t_k}
    =
    \mathbf{L}_{i,t_k}
    \mathbf{L}_{i,t_k}^{\top}
    +
    \epsilon_{\mathrm{geo}}\mathbf{I},
\end{equation}
where $\epsilon_{\mathrm{geo}}>0$ is the same small constant as in Eq.~\eqref{eq:covariance_target}.

The pair $\left(\widehat{\mathbf{x}}_{i,t_k}, \mathbf{C}_{i,t_k}\right)$ defines an anisotropic ellipsoid centered at the corrected coarse particle
position; its level set can be written as
\begin{equation}
\label{eq:ellipsoid}
    \mathcal{E}_{i,t_k}
    =
    \left\{
    \mathbf{y} \in \mathbb{R}^{d}
    \;\middle|\;
    \left(
    \mathbf{y}
    -
    \widehat{\mathbf{x}}_{i,t_k}
    \right)^{\top}
    \mathbf{C}_{i,t_k}^{-1}
    \left(
    \mathbf{y}
    -
    \widehat{\mathbf{x}}_{i,t_k}
    \right)
    \leq 1
    \right\}.
\end{equation}
Thus, a coarse particle is no longer treated as an isotropic point sample.
Instead, it acts as a structured latent element that summarizes an anisotropic
region of the reference flow.

This representation is particularly useful when $N_c < N_r$. Rather than
increasing the number of particles, the model increases the expressive capacity
of each coarse particle. The corrected center
$\widehat{\mathbf{x}}_{i,t_k}$ and velocity $\widehat{\mathbf{v}}_{i,t_k}$
capture the local dynamics, while $\mathbf{C}_{i,t_k}$ captures the local
geometry encoded by the coarse-supported reference covariance
$\mathbf{C}^{*}_{i,t_k}$. The resulting model therefore preserves the
computationally light coarse particle set while providing a richer description
of the fluid structure.

\subsection{Training Objective}
\label{sec:training_objective}

The model is trained with supervision constructed from the coarse-supported
reference targets
$\mathcal{Y}^{*}_{t_k}$ defined in
Section~\ref{sec:coarse_supported_reference_alignment}. The objective contains
two groups of constraints: dynamics constraints on the corrected coarse state
and geometry constraints on the anisotropic ellipsoidal representation.

The corrected state
$\widehat{\mathbf{x}}_{i,t_k}$ and
$\widehat{\mathbf{v}}_{i,t_k}$ is defined as in
Eq.~\eqref{eq:corrected_state}. We supervise it with position and
velocity correction losses:
\begin{equation}
\label{eq:loss_x}
    \mathcal{L}_{x}
    =
    \frac{1}{TN_f^{c}}
    \sum_{k=0}^{T-1}
    \sum_{i\in\mathcal{F}^{c}}
    \left\|
    \widehat{\mathbf{x}}_{i,t_k}
    -
    \mathbf{x}^{*}_{i,t_k}
    \right\|_{2}^{2},
\end{equation}
\begin{equation}
\label{eq:loss_v}
    \mathcal{L}_{v}
    =
    \frac{1}{TN_f^{c}}
    \sum_{k=0}^{T-1}
    \sum_{i\in\mathcal{F}^{c}}
    \left\|
    \widehat{\mathbf{v}}_{i,t_k}
    -
    \mathbf{v}^{*}_{i,t_k}
    \right\|_{2}^{2},
\end{equation}
where $N_f^{c}=|\mathcal{F}^{c}|$ is the number of coarse fluid particles.

To compare kinetic behavior across different particle resolutions, we use a
fluid-only mass-normalized kinetic energy. The corrected coarse specific kinetic energy is:
\begin{equation}
\label{eq:kinetic_coarse}
    \widehat{e}_{\mathrm{kin}}^{c}(t_k)
    =
    \frac{
    \sum_{i \in \mathcal{F}^{c}}
    \frac{1}{2}
    m^{c}_{i}
    \left\|
    \widehat{\mathbf{v}}_{i,t_k}
    \right\|_{2}^{2}
    }{
    \sum_{i \in \mathcal{F}^{c}} m^{c}_{i}
    },
\end{equation}
and the reference specific kinetic energy is:
\begin{equation}
\label{eq:kinetic_ref}
    e_{\mathrm{kin}}^{r}(t_k)
    =
    \frac{
    \sum_{j \in \mathcal{F}^{r}}
    \frac{1}{2}
    m^{r}_{j}
    \left\|
    \mathbf{v}^{r}_{j,t_k}
    \right\|_{2}^{2}
    }{
    \sum_{j \in \mathcal{F}^{r}} m^{r}_{j}
    }.
\end{equation}
In practice, a small constant is added to each denominator for numerical robustness against vanishing total mass.
The kinetic-energy auxiliary loss is then:
\begin{equation}
\label{eq:loss_kin}
    \mathcal{L}_{\mathrm{ekin}}
    =
    \frac{1}{T}
    \sum_{k=0}^{T-1}
    \left(
    \widehat{e}_{\mathrm{kin}}^{c}(t_k)
    -
    e_{\mathrm{kin}}^{r}(t_k)
    \right)^{2}.
\end{equation}

For the anisotropic ellipsoidal representation, we compare the predicted
covariance $\mathbf{C}_{i,t_k}$ with the coarse-supported reference covariance
$\mathbf{C}^{*}_{i,t_k}$. Since both matrices are symmetric positive definite,
we use a log-covariance geometry loss:
\begin{equation}
\label{eq:loss_geo}
    \mathcal{L}_{\mathrm{geo}}
    =
    \frac{1}{TN_f^{c}}
    \sum_{k=0}^{T-1}
    \sum_{i\in\mathcal{F}^{c}}
    \left\|
    \log \mathbf{C}_{i,t_k}
    -
    \log \mathbf{C}^{*}_{i,t_k}
    \right\|_{F}^{2}.
\end{equation}
Here $\log(\cdot)$ is the matrix logarithm and
$\|\cdot\|_{F}$ is the Frobenius norm. The final training objective is:

\begin{equation}
\label{eq:final_objective}
    \mathcal{L}
    =
    \lambda_{x}\mathcal{L}_{x}
    +
    \lambda_{v}\mathcal{L}_{v}
    +
    \lambda_{\mathrm{ekin}}\mathcal{L}_{\mathrm{ekin}}
    +
    \lambda_{\mathrm{geo}}\mathcal{L}_{\mathrm{geo}},
\end{equation}
where the $\lambda$ coefficients control the relative strength of the dynamics
and geometry constraints. For clarity, we write the loss for a single trajectory, in practice it is averaged over all training trajectories. For more details of the implementation of our methods, please refer to Appendix Algorithm \ref{alg:slc_training} for training and Algorithm \ref{alg:slc_inference} for inference.

\begin{algorithm}
\caption{Training Structured Latent Closure}
\label{alg:slc_training}
\KwIn{
Precomputed paired sequences
$\{(\mathcal{S}^{c}_{t_k},\mathcal{S}^{r}_{t_k})\}_{k=0}^{T-1}$,
training epochs $E$, model parameters $\theta$
}
\KwOut{Best model parameters $\theta^\star$ selected on the validation split, residual normalization statistics $(\boldsymbol{\mu}_{\Delta x},\boldsymbol{\sigma}_{\Delta x}),(\boldsymbol{\mu}_{\Delta v},\boldsymbol{\sigma}_{\Delta v})$}

\BlankLine
\textbf{Preprocessing:}\\
\ForEach{paired sequence}{
    Align $\mathcal{S}^{r}$ to the coarse time grid $\{t_k\}$\;
    \ForEach{coarse frame $t_k$ and coarse fluid particle $i\in\mathcal{F}^{c}$}{
        Collect nearby reference fluid particles around $\mathbf{x}^{c}_{i,t_k}$\;
        Compute local targets $(\mathbf{x}^{*}_{i,t_k},\mathbf{v}^{*}_{i,t_k},\mathbf{C}^{*}_{i,t_k})$\;
        Set residual targets
        $\Delta\mathbf{x}^{*}_{i,t_k}=\mathbf{x}^{*}_{i,t_k}-\mathbf{x}^{c}_{i,t_k}$ and
        $\Delta\mathbf{v}^{*}_{i,t_k}=\mathbf{v}^{*}_{i,t_k}-\mathbf{v}^{c}_{i,t_k}$\;
    }
}

\BlankLine
Estimate per-component normalization statistics $(\boldsymbol{\mu}_{\Delta x},\boldsymbol{\sigma}_{\Delta x}),(\boldsymbol{\mu}_{\Delta v},\boldsymbol{\sigma}_{\Delta v})$ from $\{\Delta\mathbf{x}^{*}_{i,t_k},\Delta\mathbf{v}^{*}_{i,t_k}\}$ on the training split\;
Initialize $\theta$, optimizer state, and $s^\star\leftarrow+\infty$ (lower is better; e.g., total validation loss)\;

\BlankLine
\For{$e=1$ \KwTo $E$}{
    \ForEach{mini-batch $\mathcal{B}$ of coarse frames}{
        For each coarse fluid particle $i\in\mathcal{F}^{c}$, extract learned features and predict closure:
        \[
        \mathbf{h}_{i,t_k} = \phi_{\theta}(\mathcal{S}^{c}_{t_k}, i),
        \]
        \[
        \mathbf{C}_{i,t_k} = g_{\theta}^{\mathrm{geo}}(\mathbf{h}_{i,t_k}),
        \quad
        (\bar{\Delta\mathbf{x}}_{i,t_k},\bar{\Delta\mathbf{v}}_{i,t_k}) = g_{\theta}^{\mathrm{res}}(\mathbf{h}_{i,t_k}).
        \]

        Denormalize residuals $(\Delta\mathbf{x}_{i,t_k},\Delta\mathbf{v}_{i,t_k})$ via Eq.~\eqref{eq:denormalize} and obtain the corrected coarse state $(\widehat{\mathbf{x}}_{i,t_k},\widehat{\mathbf{v}}_{i,t_k})$ via Eq.~\eqref{eq:corrected_state}\;

        Compute
        $\mathcal{L}_{x}$,
        $\mathcal{L}_{v}$,
        $\mathcal{L}_{\mathrm{ekin}}$,
        and
        $\mathcal{L}_{\mathrm{geo}}$
        via Eqs.~\eqref{eq:loss_x},~\eqref{eq:loss_v},~\eqref{eq:loss_kin},~\eqref{eq:loss_geo}\;

        Update $\theta$ by minimizing $\mathcal{L}$ (Eq.~\eqref{eq:final_objective})\;
    }
    Evaluate the validation split and compute selection score $s_e$\;
    \If{$s_e < s^\star$}{
        $s^\star\leftarrow s_e$ and $\theta^\star\leftarrow\theta$\;
    }
}

\Return{$\theta^\star$, $(\boldsymbol{\mu}_{\Delta x},\boldsymbol{\sigma}_{\Delta x}),(\boldsymbol{\mu}_{\Delta v},\boldsymbol{\sigma}_{\Delta v})$}\;
\end{algorithm}
\newpage
\begin{algorithm}[ht]
\caption{Inference with Structured Latent Closure}
\label{alg:slc_inference}
\KwIn{
A coarse sequence
$\{\mathcal{S}^{c}_{t_k}\}_{k=0}^{T-1}$,
trained parameters $\theta^\star$,
residual normalization statistics
$(\boldsymbol{\mu}_{\Delta x},\boldsymbol{\sigma}_{\Delta x})$
and
$(\boldsymbol{\mu}_{\Delta v},\boldsymbol{\sigma}_{\Delta v})$
}
\KwOut{
Corrected coarse sequence
$\{\widehat{\mathcal{S}}_{t_k}\}_{k=0}^{T-1}$, anisotropic geometry
$\{\mathbf{C}_{i,t_k}\}_{i\in\mathcal{F}^{c},\,k=0,\dots,T-1}$
}

\BlankLine
\ForEach{coarse frame $t_k$}{
    \ForEach{coarse fluid particle $i\in\mathcal{F}^{c}$}{
        Extract learned features and predict closure:
        \[
        \mathbf{h}_{i,t_k} = \phi_{\theta^\star}(\mathcal{S}^{c}_{t_k}, i),
        \]
        \[
        \mathbf{C}_{i,t_k} = g_{\theta^\star}^{\mathrm{geo}}(\mathbf{h}_{i,t_k}),
        \quad
        (\bar{\Delta\mathbf{x}}_{i,t_k},\bar{\Delta\mathbf{v}}_{i,t_k}) = g_{\theta^\star}^{\mathrm{res}}(\mathbf{h}_{i,t_k}).
        \]

        Denormalize residuals $(\Delta\mathbf{x}_{i,t_k},\Delta\mathbf{v}_{i,t_k})$ via Eq.~\eqref{eq:denormalize} and obtain the corrected coarse state $(\widehat{\mathbf{x}}_{i,t_k},\widehat{\mathbf{v}}_{i,t_k})$ via Eq.~\eqref{eq:corrected_state}\;
    }
    Pass non-fluid particles ($i\notin\mathcal{F}^{c}$) through unchanged: $\widehat{\mathbf{x}}_{i,t_k}=\mathbf{x}^{c}_{i,t_k}$, $\widehat{\mathbf{v}}_{i,t_k}=\mathbf{v}^{c}_{i,t_k}$\;

    Form the corrected frame by replacing positions and velocities in $\mathcal{S}^{c}_{t_k}$ with corrected values:
    \[
    \widehat{\mathcal{S}}_{t_k}
    =
    \{
    (\widehat{\mathbf{x}}_{i,t_k},
     \widehat{\mathbf{v}}_{i,t_k},
     \rho^{c}_{i,t_k},
     p^{c}_{i,t_k},
     m^{c}_{i},
     \tau^{c}_{i})
    \}_{i=1}^{N_c}.
    \]
}

\Return{
$\{\widehat{\mathcal{S}}_{t_k}\}_{k=0}^{T-1}$
and
$\{\mathbf{C}_{i,t_k}\}_{i\in\mathcal{F}^{c},\,k=0,\dots,T-1}$
}\;
\end{algorithm}

%% file: sections/experiments.tex
\section{Experiments}

\subsection{Experimental Setup and implementation}
\label{sec:experimental_setup}
\begin{table}[ht!]
    \centering
    \caption{
    Dataset statistics for paired particle trajectories. Particle counts are reported as fluid/total. Split sizes denote the number of paired trajectories in train/validation/test, and frames denote saved frames per trajectory. Here $N_c$ and $N_r$ denote the number of coarse and reference particles, respectively, and $T$ denotes the number of saved frames per trajectory.
    }
    \label{tab:dataset_statistics}
    \small
    \setlength{\tabcolsep}{6pt}
    \renewcommand{\arraystretch}{1.08}
    \begin{tabular}{lcccc}
        \toprule
        Dataset
        & $N_c$ 
        & $N_r$
        & Train/Val/Test
        & Frames/traj. \\
        \midrule
        2D TGV
        & $484/484$
        & $2500/2500$
        & $100/50/50$
        & $126$ \\

        2D DAM
        & $1250/1622$
        & $5000/5740$
        & $50/25/25$
        & $401$ \\

        2D LDC
        & $500/580$
        & $2500/2708$
        & $1/1/1$
        & $10001$ \\

        2D RPF
        & $800/800$
        & $3200/3200$
        & $1/1/1$
        & $10001$ \\

        3D TGV
        & $1000/1000$
        & $8000/8000$
        & $200/100/100$
        & $61$ \\

        3D LDC
        & $2048/2608$
        & $6912/8160$
        & $1/1/1$
        & $5001$ \\

        3D RPF
        & $1000/1000$
        & $8000/8000$
        & $1/1/1$
        & $5001$ \\
        \bottomrule
    \end{tabular}
\end{table}
We evaluate our method on paired coarse and high-resolution reference
sequences. Each dataset is split into training, validation, and test subsets as
summarized in Table~\ref{tab:dataset_statistics}. The model is trained on the
training split, the best checkpoint is selected on the validation split, and
final results are reported on the held-out test split. All experiments are
conducted on a machine with 2 Nvidia H200 GPUs.

All models use a two-layer MLP with hidden dimension $64$ to predict the current-frame residuals
$(\Delta \mathbf{x}_{i,t}, \Delta \mathbf{v}_{i,t})$ and the anisotropic covariance
$\mathbf{C}_{i,t}$ from coarse particle features. We train with Adam and select the checkpoint
with the lowest validation composite score. Unless otherwise specified, we use learning rate
$3\times 10^{-4}$, batch size $32$, gradient clipping with norm $1.0$, and loss weights
$\lambda_{\Delta x}=\lambda_{\Delta v}=2.0$, $\lambda_{\mathrm{geo}}=1.0$,
$\lambda_{\mathrm{dyn}}=1.0$, and $\lambda_{\mathrm{ekin}}=0.5$.
For the more unstable 3D TGV setting, we use a smaller learning rate $1\times 10^{-4}$ and
gradient clipping norm $0.5$. As shown in the Table \ref{tab:training_hyperparams}.

\begin{table}[t]
    \centering
    \caption{Training hyperparameters used in the main experiments. All runs use Adam, hidden dimension $64$, and validation-based checkpoint selection.}
    \label{tab:training_hyperparams}
    \small
    \setlength{\tabcolsep}{5pt}
    \renewcommand{\arraystretch}{1.08}
    \begin{tabular}{llcccc}
        \toprule
        Solver & Dataset
        & Epochs
        & Batch size
        & Learning rate
        & Grad. clip \\
        \midrule
        SPH & 2D TGV / 2D DAM / 2D LDC / 2D RPF
        & 30 & 32 & $3.0\mathrm{e}{-4}$ & 1.0 \\
        SPH & 3D TGV
        & 40 & 32 & $1.0\mathrm{e}{-4}$ & 0.5 \\
        SPH & 3D LDC
        & 30 & 32 & $3.0\mathrm{e}{-4}$ & 1.0 \\
        SPH & 3D RPF
        & 10 & 32 & $2.0\mathrm{e}{-4}$ & 0.5 \\
        \midrule
        MPS & 2D DAM / 2D LDC / 2D RPF
        & 30 & 32 & $3.0\mathrm{e}{-4}$ & 1.0 \\
        MPS & 2D TGV
        & 40 & 32 & $3.0\mathrm{e}{-4}$ & 1.0 \\
        MPS & 3D TGV
        & 40 & 32 & $1.0\mathrm{e}{-4}$ & 0.5 \\
        MPS & 3D LDC / 3D RPF
        & 30 & 32 & $3.0\mathrm{e}{-4}$ & 1.0 \\
        \bottomrule
    \end{tabular}
\end{table}

\subsection{Results}
\begin{figure}
\label{fig:uniform_db2d}
    \centering
    \includegraphics[width=\linewidth]{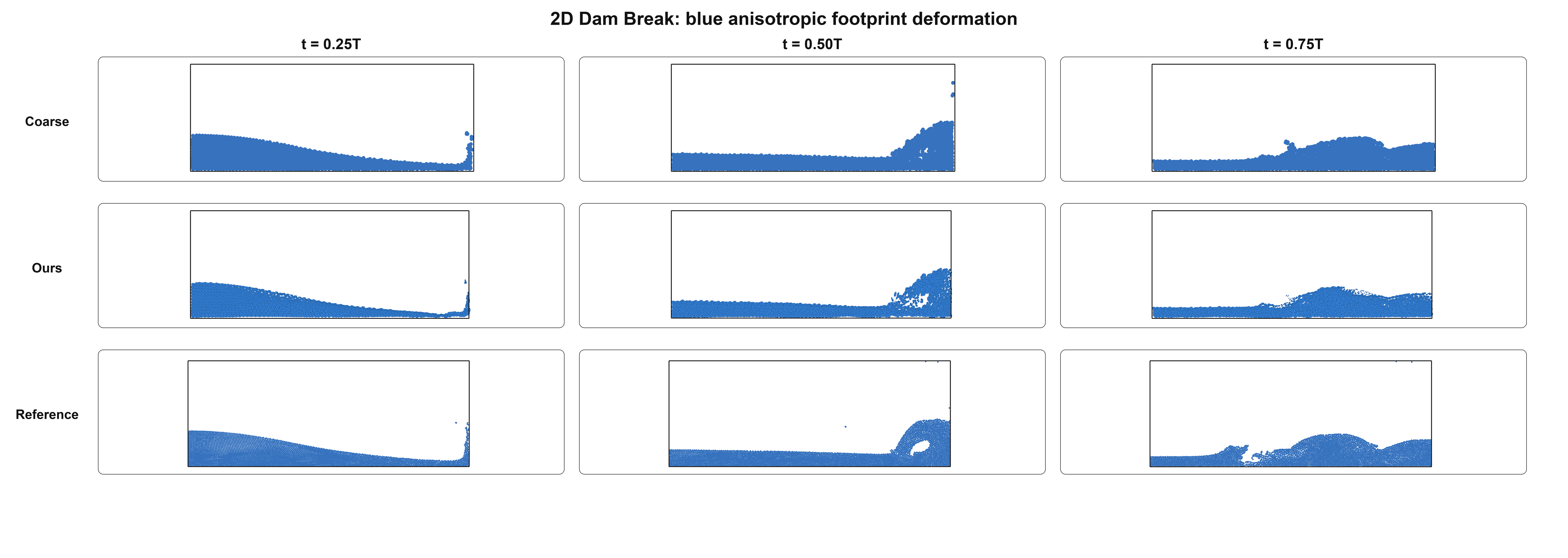}
    \caption{
Qualitative visualization on the 2D dam-break scene using a uniform blue rendering. 
The coarse baseline represents each particle as an isotropic point, while our method augments the same coarse particles with learned anisotropic ellipsoidal footprints. 
In regions with richer motion, our representation learns visibly elongated particle supports, indicating stronger local structural expressiveness under the same particle count. 
At $t=0.50T$, near the liquid fallback region, our corrected state preserves a more coherent structure than the coarse particles. 
At $t=0.75T$, our method also avoids the abnormal wave crest observed in the coarse baseline, producing a shape closer to the high-resolution reference without particle densification.
}
\end{figure}

We report mean squared error (MSE) in position, velocity and the mass-normalized specific kinetic-energy error. Each sample is a
saved frame \(\left(\mathcal{S}^{c}_{t_k}, \mathcal{S}^{r}_{t_k}\right)\).
The model takes only the coarse state \(\mathcal{S}^{c}_{t_k}\) as input and
predicts current-frame residual corrections for the coarse particles. The
reference state \(\mathcal{S}^{r}_{t_k}\) is used only to construct
coarse-supported supervision targets \(\mathcal{Y}^{*}_{t_k}\) and to compute
evaluation metrics.
Our experiments are designed to answer three questions:
(i) whether AnisoLift improves coarse particle fidelity under a fixed particle budget;
(ii) whether anisotropic geometry provides useful local structural information beyond state residuals;
and (iii) whether the additional inference stage preserves the efficiency advantage. We show quantitative results in the Table \ref{tab:main_results_all}, where we can see our method improve the fidelity of coarse particles generated by SPH and MPS. We also show qualitative results on dataset synthesized by SPH. In the Figure \ref{fig:uniform_db2d}, we show the learned structure given by our method in the middle row on 2D DB, ellipsoidal shape can be observed in each particle, and our method shows cavities appearing when water waves recede in $t=0.50T$, while coarse particles failed to construct such structures. 

\begin{figure}[ht]
    \centering
    \includegraphics[width=\linewidth]{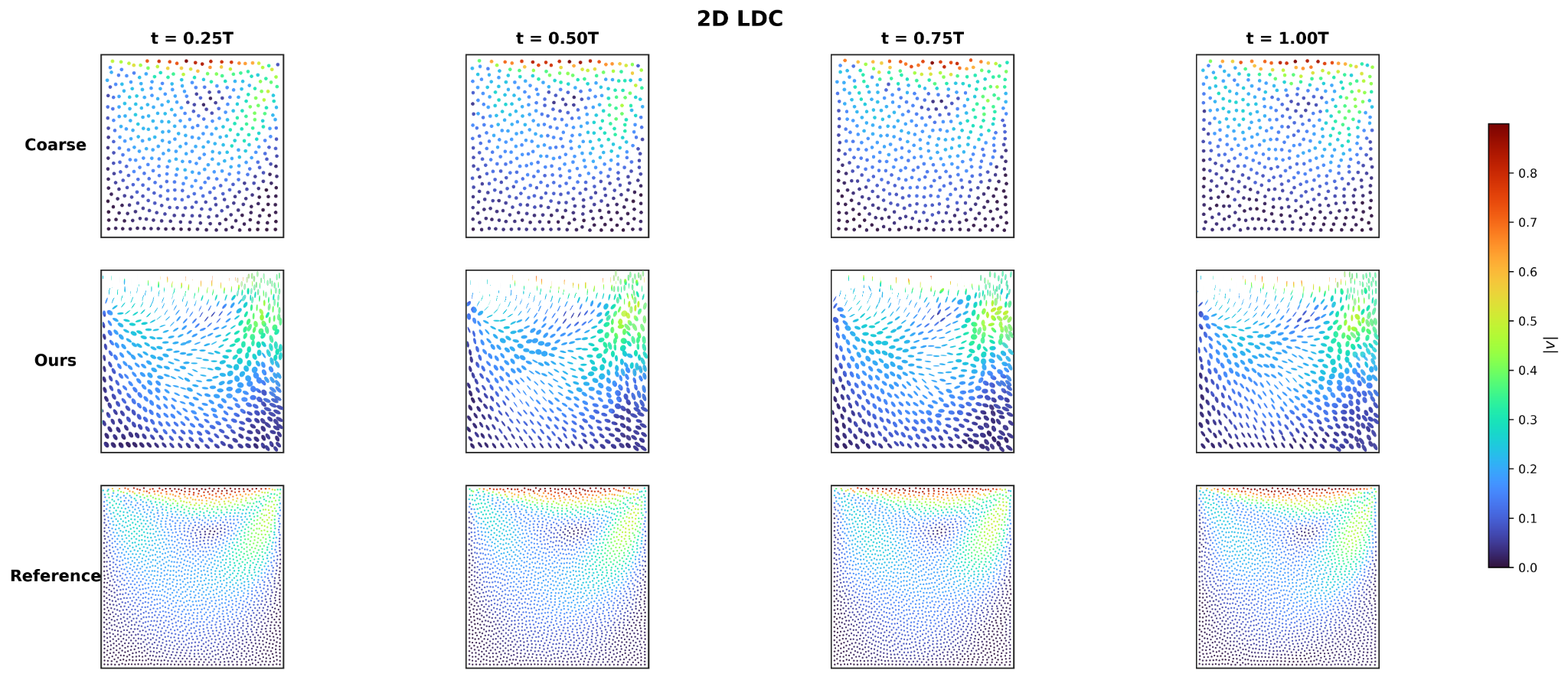}
    \caption{Qualitative correction on the 2D lid-driven cavity dataset. Columns show four normalized time instants, and rows compare the coarse solver trajectory, AnisoLift, and the reference. Particles are colored by velocity magnitude. AnisoLift preserves the same coarse particle count but replaces point particles with learned anisotropic ellipsoidal footprints, revealing locally directional flow structure and a smoother cavity circulation pattern that is not captured by the coarse point representation.}
    \label{fig:ldc2d_footprints_velocity_ellipses_recolored_ours_match_reference}
\end{figure}


\subsection{Evaluation metrics}
We evaluate the corrected coarse state against the coarse-supported reference
targets and the original high-resolution reference sequence. Since the coarse
and reference have different particle counts, pointwise state
errors are computed on aligned coarse support. For position and velocity, we report
\begin{equation}
    \mathrm{MSE}_{x}
    =
    \frac{1}{TN_f^c}
    \sum_{k=0}^{T-1}
    \sum_{i\in\mathcal{F}^c}
    \left\|
    \widehat{\mathbf{x}}_{i,t_k}
    -
    \mathbf{x}^{*}_{i,t_k}
    \right\|_{2}^{2},~ 
    \mathrm{MSE}_{v}
    =
    \frac{1}{TN_f^c}
    \sum_{k=0}^{T-1}
    \sum_{i\in\mathcal{F}^c}
    \left\|
    \widehat{\mathbf{v}}_{i,t_k}
    -
    \mathbf{v}^{*}_{i,t_k}
    \right\|_{2}^{2}.
\end{equation}
For the uncorrected coarse state, we replace
$\widehat{\mathbf{x}}_{i,t_k}$ and $\widehat{\mathbf{v}}_{i,t_k}$ with
$\mathbf{x}^{c}_{i,t_k}$ and $\mathbf{v}^{c}_{i,t_k}$.

To evaluate global kinetic behavior under different particle counts, we use
fluid-only mass-normalized specific kinetic energy. For any particle state
$\mathcal{S}_{t_k}$ with fluid particle set $\mathcal{F}$, we define
\begin{equation}
    e_{\mathrm{kin}}(t_k)
    =
    \frac{
    \sum_{i \in \mathcal{F}}
    \frac{1}{2} m_i
    \left\|
    \mathbf{v}_{i,t_k}
    \right\|_{2}^{2}
    }{
    \sum_{i \in \mathcal{F}} m_i + \epsilon
    }.
\end{equation}
Here $\epsilon$ is a small positive constant for numerical stability.
The specific kinetic-energy error is
\begin{equation}
    \mathrm{MSE}_{e_{\mathrm{kin}}}
    =
    \frac{1}{T}
    \sum_{k=0}^{T-1}
    \left(
    \widehat{e}_{\mathrm{kin}}^{c}(t_k)
    -
    e_{\mathrm{kin}}^{r}(t_k)
    \right)^{2},
\end{equation}
where $\widehat{e}_{\mathrm{kin}}^{c}(t_k)$ is computed from the corrected
coarse state and $e_{\mathrm{kin}}^{r}(t_k)$ is computed from the
high-resolution reference state. This metric avoids directly comparing raw
total kinetic energy across particle sets with different resolutions.

\subsection{Datasets}
\label{sec:datasets}

We evaluate on multiple particle-based liquid simulation scenarios in both 2D
and 3D, including Taylor--Green vortex (TGV), dam break (DAM), lid-driven cavity
(LDC), and reverse Poiseuille flow (RPF). For each scenario, we construct paired
coarse and high-resolution reference sequences with matched physical duration,
output interval, and number of saved frames. The coarse sequence uses fewer
particles and provides the low-resolution input, while the high-resolution
sequence provides supervision and evaluation targets. More details are shown in the Table \ref{tab:dataset_statistics}.


\begin{table}[h]
    \centering
    \caption{Quantitative results on 2D and 3D paired particle simulation datasets. Ekin denotes fluid-only mass-normalized specific kinetic energy. The scale in each dataset header applies to all three metrics. Smaller value reflects better performance.}
    \label{tab:main_results_all}
    \small
    \setlength{\tabcolsep}{3pt}
    \renewcommand{\arraystretch}{1.08}

    \begin{tabular*}{\textwidth}{@{\extracolsep{\fill}}lcccccccccccc}
        \toprule
        \multirow{2}{*}{Solver}
        & \multicolumn{3}{c}{2D TGV (1e-5)}
        & \multicolumn{3}{c}{2D DAM (1e-3)}
        & \multicolumn{3}{c}{2D LDC (1e-3)}
        & \multicolumn{3}{c}{2D RPF (1e-3)} \\
        \cmidrule(lr){2-4}
        \cmidrule(lr){5-7}
        \cmidrule(lr){8-10}
        \cmidrule(lr){11-13}
        & {\tiny$\mathrm{MSE}_x$} & {\tiny$\mathrm{MSE}_v$} & {\tiny$\mathrm{MSE}_{\text{Ekin}}$}
        & {\tiny$\mathrm{MSE}_x$} & {\tiny$\mathrm{MSE}_v$} & {\tiny$\mathrm{MSE}_{\text{Ekin}}$}
        & {\tiny$\mathrm{MSE}_x$} & {\tiny$\mathrm{MSE}_v$} & {\tiny$\mathrm{MSE}_{\text{Ekin}}$}
        & {\tiny$\mathrm{MSE}_x$} & {\tiny$\mathrm{MSE}_v$} & {\tiny$\mathrm{MSE}_{\text{Ekin}}$} \\
        \midrule
        $\mathrm{SPH}_{c}$
        & 7.750 & 256.46 & 2.360
        & 2.560 & 35.92 & 0.250
        & 0.6420 & 21.79 & 0.010
        & 0.0800 & 8.200 & 2.350 \\

        $\mathrm{SPH}_{\mathrm{Ours}}$
        & 2.010 & 221.60 & 2.350
        & 1.100 & 22.34 & 0.110
        & 0.0020 & 2.690 & 0.003
        & 0.0080 & 1.690 & 0.170 \\
        \midrule
        $\mathrm{MPS}_{c}$
        & 12.70 & 460.78 & 387.20
        & 7.330 & 710.27 & 32.40
        & 7.050 & 203.70 & 1.480
        & 0.1200 & 7.950 & 61.51 \\

        $\mathrm{MPS}_{\mathrm{Ours}}$
        & 7.600 & 367.84 & 298.66
        & 3.710 & 368.99 & 28.19
        & 1.510 & 91.70 & 0.480
        & 0.0300 & 1.580 & 8.300 \\

    \end{tabular*}

    \vspace{0.8em}

    \begin{tabular*}{\textwidth}{@{\extracolsep{\fill}}lccccccccc}
        \toprule
        \multirow{2}{*}{Solver}
        & \multicolumn{3}{c}{3D TGV (1e-3)}
        & \multicolumn{3}{c}{3D LDC (1e-3)}
        & \multicolumn{3}{c}{3D RPF (1e-3)} \\
        \cmidrule(lr){2-4}
        \cmidrule(lr){5-7}
        \cmidrule(lr){8-10}
        & {\tiny$\mathrm{MSE}_x$} & {\tiny$\mathrm{MSE}_v$} & {\tiny$\mathrm{MSE}_{\text{Ekin}}$}
        & {\tiny$\mathrm{MSE}_x$} & {\tiny$\mathrm{MSE}_v$} & {\tiny$\mathrm{MSE}_{\text{Ekin}}$}
        & {\tiny$\mathrm{MSE}_x$} & {\tiny$\mathrm{MSE}_v$} & {\tiny$\mathrm{MSE}_{\text{Ekin}}$} \\
        \midrule
        $\mathrm{SPH}_{c}$
        & 0.0519 & 21.26 & 0.0397
        & 0.32 & 3.16 & 0.009
        & 4.54 & 8.99 & 2.90 \\

        $\mathrm{SPH}_{\mathrm{Ours}}$
        & 0.0517 & 12.54 & 0.0012
        & 0.03 & 0.49 & 0.008
        & 2.87 & 2.15 & 1.88 \\
        \midrule
        $\mathrm{MPS}_{c}$
        & 15.84 & 79.64 & 48.15
        & 0.50 & 9.92 & 0.110
        & 1.88 & 9.00 & 3.11 \\

        $\mathrm{MPS}_{\mathrm{Ours}}$
        & 5.43 & 39.82 & 32.53
        & 0.41 & 4.89 & 0.004
        & 0.08 & 2.85 & 1.22 \\
        \bottomrule
    \end{tabular*}
\end{table}

The SPH-generated datasets are used for the main evaluation, while the
MPS-generated datasets are used to assess solver generalization. In all cases,
the model operates only on the coarse particles and does not increase the
particle count at inference time.

\paragraph{Paired trajectory generation.}
For each benchmark, we generate paired coarse and high-resolution reference trajectories using
the same solver family and the same saved temporal resolution. The coarse trajectory is produced
by a stable solver rollout and is used directly during training; no differentiable solver rollout
is used in the learning stage. We evaluate two particle solvers: SPH and MPS. The SPH datasets are
generated with JAX-SPH, while the MPS datasets are generated with MPS-Basic using an OpenMP CPU
backend. For MPS-Basic, we use implicit pressure solve, collision distance ratio $0.5$, and quality
checks that reject trajectories containing non-finite positions, velocities, densities, or pressures.
For periodic cases, SPH uses native periodic boundary handling; MPS-Basic uses the closest available
bounded or periodic-gauge approximation depending on the case.

\paragraph{Particle solvers.}
We use two classical particle solvers to generate paired simulation data:
smoothed particle hydrodynamics (SPH) and the moving particle semi-implicit
method (MPS). SPH represents the fluid as Lagrangian particles and estimates
field quantities through kernel interpolation over neighboring particles. A
generic SPH approximation of a scalar field $a$ at particle $i$ can be written
as
\begin{equation}
    a_i
    \approx
    \sum_{j}
    \frac{m_j}{\rho_j}
    a_j
    W
    \left(
    \left\|
    \mathbf{x}_i - \mathbf{x}_j
    \right\|, h
    \right),
\end{equation}
where $W(\cdot,h)$ is a smoothing kernel with support radius controlled by
$h$. In our experiments, SPH provides stable low- and high-resolution particle
rollouts with matched output times.

MPS is also a fully Lagrangian particle method, but it does not rely on the
standard SPH density-kernel formulation. Instead, it computes particle number
density from neighbor weights and solves pressure implicitly to enforce
near-incompressibility. A typical MPS number density estimate is
\begin{equation}
    n_i
    =
    \sum_{j \neq i}
    w
    \left(
    \left\|
    \mathbf{x}_j - \mathbf{x}_i
    \right\|
    \right),
\end{equation}
where $w(\cdot)$ is the MPS interaction weight. We include MPS-generated
datasets to test whether the proposed closure is tied to SPH-specific
discretization artifacts or can generalize to particle trajectories produced
by a different solver family.

\subsection{Ablation study}

\paragraph{Loss-component ablation on 2D LDC.}
We ablate the main loss terms on the 2D lid-driven cavity dataset using both SPH- and MPS-generated paired trajectories. The coarse baseline has no learned correction. For learned variants, we remove one loss group at a time: corrected-state losses, residual-consistency losses, log-covariance geometry loss, and specific kinetic-energy loss.

Table~\ref{tab:ablation_ldc2d_loss} shows that all terms are necessary but influence different aspects. Removing corrected-state losses degrades position and velocity accuracy, while removing residual-consistency losses reduces correction quality. The geometry loss is essential for learning meaningful anisotropic structure, and the kinetic-energy term regularizes global behavior, preventing locally accurate but physically inconsistent updates.
Overall, the results support combining state supervision, residual consistency, geometric supervision, and a fluid-only kinetic-energy regularizer. These components are complementary: state and residual terms drive local accuracy, geometry shapes anisotropic representation, and the kinetic term improves global consistency.
\begin{table}[h]
    \centering
    \caption{Loss-component ablation on the 2D LDC dataset under SPH- and MPS-generated paired trajectories. The coarse baseline is reported first, followed by ablated variants. Ekin denotes fluid-only mass-normalized specific kinetic energy. Lower values are better.}
    \label{tab:ablation_ldc2d_loss}
    \small
    \setlength{\tabcolsep}{5pt}
    \renewcommand{\arraystretch}{1.08}
    \begin{tabular}{lcccccc}
        \toprule
        \multirow{2}{*}{Variant}
        & \multicolumn{3}{c}{SPH}
        & \multicolumn{3}{c}{MPS} \\
        \cmidrule(lr){2-4}
        \cmidrule(lr){5-7}
        & $\mathrm{MSE}_{x}$
        & $\mathrm{MSE}_{v}$
        & $\mathrm{MSE}_{\mathrm{Ekin}}$
        & $\mathrm{MSE}_{x}$
        & $\mathrm{MSE}_{v}$
        & $\mathrm{MSE}_{\mathrm{Ekin}}$ \\
        \midrule
        Coarse
        & $6.352\mathrm{e}{-4}$
        & $2.179\mathrm{e}{-2}$
        & $1.339\mathrm{e}{-5}$
        & $7.051\mathrm{e}{-3}$
        & $2.037\mathrm{e}{-1}$
        & $1.478\mathrm{e}{-3}$ \\
        
        w/o Ekin
        & $2.802\mathrm{e}{-4}$
        & $1.104\mathrm{e}{-2}$
        & $8.524\mathrm{e}{-5}$
        & $1.734\mathrm{e}{-3}$
        & $1.112\mathrm{e}{-1}$
        & $2.883\mathrm{e}{-3}$ \\

        w/o Geometry
        & $6.895\mathrm{e}{-6}$
        & $2.930\mathrm{e}{-3}$
        & $2.147\mathrm{e}{-6}$
        & \textbf{$8.622\mathrm{e}{-4}$}
        & \textbf{$6.270\mathrm{e}{-2}$}
        & $9.087\mathrm{e}{-4}$ \\

        w/o Residual
        & $3.210\mathrm{e}{-6}$
        & $2.984\mathrm{e}{-3}$
        & $1.964\mathrm{e}{-6}$
        & $1.506\mathrm{e}{-3}$
        & $9.294\mathrm{e}{-2}$
        & $3.762\mathrm{e}{-4}$ \\

        w/o State
        & $5.238\mathrm{e}{-6}$
        & $3.614\mathrm{e}{-3}$
        & \textbf{$1.351\mathrm{e}{-6}$}
        & $1.529\mathrm{e}{-3}$
        & $9.481\mathrm{e}{-2}$
        & \textbf{$1.611\mathrm{e}{-4}$} \\
        \midrule
        Full loss
        & \textbf{$2.102\mathrm{e}{-6}$}
        & \textbf{$2.659\mathrm{e}{-3}$}
        & $3.131\mathrm{e}{-6}$
        & $1.506\mathrm{e}{-3}$
        & $9.170\mathrm{e}{-2}$
        & $4.834\mathrm{e}{-4}$ \\
        \bottomrule
    \end{tabular}
\end{table}

\paragraph{Resolution scaling on 2D TGV.}
We evaluate the proposed correction under different coarse particle budgets on the 2D Taylor--Green vortex dataset. The high-resolution reference is fixed at \(N_r=2500\), while coarse trajectories are generated with varying \(N_c\). This tests whether the structured latent closure remains effective when the coarse simulation is either highly sparse or moderately resolved.

Table~\ref{tab:tgv2d_resolution_scaling} reports results for \(N_c=484\) and \(N_c=1225\). In both cases, our method reduces the position error relative to the coarse baseline, indicating that the learned current-frame residual correction consistently improves spatial fidelity across resolutions. The gain is larger at lower resolution, where the coarse transport skeleton misses more local structure, and smaller but still positive at higher resolution, where the baseline is already closer to the reference.

Improvements in velocity and kinetic energy are more moderate. This is expected because the correction is applied frame-wise without differentiable solver rollout, while the coarse solver already provides a stable velocity field at the same temporal resolution. Overall, the method is most beneficial under strong coarse compression, while still yielding measurable gains at higher coarse resolutions.
\begin{table}[H]
    \centering
    \caption{Resolution scaling on 2D TGV. The high-resolution reference is fixed with $N_r=2500$. Gain denotes the relative error reduction of our method over the corresponding coarse baseline.}
    \label{tab:tgv2d_resolution_scaling}
    \small
    \setlength{\tabcolsep}{4pt}
    \renewcommand{\arraystretch}{1.08}
    \begin{tabular}{lccccccccc}
        \toprule
        \multirow{2}{*}{$N_c$}
        & \multicolumn{3}{c}{$\mathrm{MSE}_{x}$}
        & \multicolumn{3}{c}{$\mathrm{MSE}_{v}$}
        & \multicolumn{3}{c}{$\mathrm{MSE}_{\mathrm{\text{Ekin}}}$} \\
        \cmidrule(lr){2-4}
        \cmidrule(lr){5-7}
        \cmidrule(lr){8-10}
        & Coarse & Ours & Gain
        & Coarse & Ours & Gain
        & Coarse & Ours & Gain \\
        \midrule
        484
        & $7.754e{-5}$ & $2.009e{-5}$ & $74.1\%$
        & $2.565e{-3}$ & $2.216e{-3}$ & $13.6\%$
        & $2.357e{-5}$ & $2.711e{-5}$ & $15.0\%$ \\

        1225
        & $2.029e{-5}$ & $9.203e{-6}$ & $54.7\%$
        & $8.130e{-4}$ & $7.768e{-4}$ & $4.46\%$
        & $3.473e{-6}$ & $3.166e{-6}$ & $8.84\%$ \\
        \bottomrule
    \end{tabular}
\end{table}
\begin{table}[H]
    \centering
    \caption{Ablation of anisotropic versus isotropic ellipsoidal representations on 2D LDC. Ekin denotes fluid-only mass-normalized specific kinetic energy. Lower is better.}
    \label{tab:ablation_anisotropic_isotropic_ldc2d}
    \small
    \setlength{\tabcolsep}{5pt}
    \renewcommand{\arraystretch}{1.08}
    \begin{tabular}{lcccccc}
        \toprule
        \multirow{2}{*}{Variant}
        & \multicolumn{3}{c}{SPH}
        & \multicolumn{3}{c}{MPS} \\
        \cmidrule(lr){2-4}
        \cmidrule(lr){5-7}
        & $\mathrm{MSE}_{x}$
        & $\mathrm{MSE}_{v}$
        & $\mathrm{MSE}_{\mathrm{Ekin}}$
        & $\mathrm{MSE}_{x}$
        & $\mathrm{MSE}_{v}$
        & $\mathrm{MSE}_{\mathrm{Ekin}}$ \\
        \midrule
        Coarse
        & $6.352\mathrm{e}{-4}$
        & $2.179\mathrm{e}{-2}$
        & $1.339\mathrm{e}{-5}$
        & $7.051\mathrm{e}{-3}$
        & $2.037\mathrm{e}{-1}$
        & $1.478\mathrm{e}{-3}$ \\

        Anisotropic
        & $1.102\mathrm{e}{-6}$
        & $2.659\mathrm{e}{-3}$
        & $2.131\mathrm{e}{-6}$
        & $6.506\mathrm{e}{-3}$
        & $5.170\mathrm{e}{-2}$
        & $4.834\mathrm{e}{-3}$ \\

        Isotropic
        & $2.066\mathrm{e}{-6}$
        & $2.668\mathrm{e}{-3}$
        & $3.534\mathrm{e}{-6}$
        & $8.597\mathrm{e}{-4}$
        & $9.976\mathrm{e}{-2}$
        & $7.705\mathrm{e}{-4}$ \\
        \bottomrule
    \end{tabular}
\end{table}

\paragraph{Anisotropic vs. isotropic representation on 2D LDC.}
We further examine whether the anisotropic ellipsoidal representation is necessary, or whether an isotropic footprint is sufficient. In this ablation, we keep the same current-frame correction pipeline and training objective, but replace the full covariance matrix \(C_i\) with an isotropic covariance parameterized by a single scale, so that each corrected coarse particle represents a circular rather than oriented local footprint.

Table~\ref{tab:ablation_anisotropic_isotropic_ldc2d} compares the coarse baseline, the full anisotropic model, and the isotropic variant on SPH- and MPS-generated 2D LDC trajectories. The anisotropic model performs better because it can adapt both the scale and orientation of coarse particles footprint to local flow structure. This is beneficial in shear-dominated regions and wall-induced directional transport, where the reference distribution is not well captured by an isotropic neighborhood. The isotropic variant still improves over the coarse baseline through residual state correction, but its representational capacity is more limited.
Overall, the results show that anisotropic geometry provides an additional advantage by allowing each coarse particle to encode directional local support without increasing the particle count.

%% file: sections/conclusion.tex
\section{Conclusion and discussion}
\label{sec:conclusion}

We presented AnisoLift, a structured latent closure framework for enhancing coarse particle liquid simulations under a fixed particle budget. Instead of generating additional particles, our method enriches each coarse particle with current-frame residual state corrections and an anisotropic ellipsoidal latent representation. Across paired 2D and 3D particle benchmarks, this design consistently improves coarse-simulation fidelity, demonstrating that coarse particles can be made substantially more expressive without densification or differentiable solver rollout during training.

At the same time, our method has several important limitations. First, it relies on paired coarse and high-resolution trajectories, so its performance depends on the quality and coverage of the generated supervision data. Second, AnisoLift performs frame-wise correction rather than closed-loop rollout, and therefore does not directly improve the stability of the underlying coarse solver. Third, because the learned closure is local, it may be insufficient when the coarse trajectory misses important global transport behavior that cannot be recovered from local anisotropic structure alone. Besides, learned corrections may inherit biases from the particle solver used to generate the supervision data.

These observations suggest several directions for future work, including integrating structured latent closure with temporally consistent or closed-loop correction, extending the framework to broader solver families, and exploring richer local representations beyond covariance-based anisotropy. More broadly, AnisoLift highlights a practical fixed-budget alternative to particle densification by increasing the expressive capacity of existing particles.

%% file: references.bib
@inproceedings{xing2025gaussianfluids,
  author    = {Xing, Jingrui and Wang, Bin and Chu, Mengyu and Chen, Baoquan},
  title     = {Gaussian Fluids: A Grid-Free Fluid Solver based on Gaussian Spatial Representation},
  year      = {2025},
  isbn      = {9798400715402},
  publisher = {Association for Computing Machinery},
  address   = {New York, NY, USA},
  url       = {https://doi.org/10.1145/3721238.3730620},
  doi       = {10.1145/3721238.3730620},
  booktitle = {ACM SIGGRAPH 2025 Conference Papers},
  articleno = {31},
  numpages  = {11},
  location  = {Vancouver, BC, Canada},
  series    = {SIGGRAPH '25}
}

@article{monaghan1992sph,
  title   = {Smoothed Particle Hydrodynamics},
  author  = {Monaghan, J. J.},
  journal = {Annual Review of Astronomy and Astrophysics},
  volume  = {30},
  pages   = {543--574},
  year    = {1992},
  doi     = {10.1146/annurev.aa.30.090192.002551}
}

@article{monaghan2005sph,
  title   = {Smoothed particle hydrodynamics},
  author  = {Monaghan, J. J.},
  journal = {Reports on Progress in Physics},
  volume  = {68},
  number  = {8},
  pages   = {1703--1759},
  year    = {2005},
  doi     = {10.1088/0034-4885/68/8/R01}
}

@inproceedings{becker2007wcsph,
  title     = {Weakly compressible {SPH} for free surface flows},
  author    = {Becker, Markus and Teschner, Matthias},
  booktitle = {Proceedings of the 2007 ACM SIGGRAPH/Eurographics Symposium on Computer Animation},
  year      = {2007},
  doi       = {10.1145/1272690.1272719},
  url       = {https://dl.acm.org/doi/10.5555/1272690.1272719}
}

@inproceedings{solenthaler2009pcisph,
author = {Solenthaler, B. and Pajarola, R.},
title = {Predictive-corrective incompressible SPH},
year = {2009},
isbn = {9781605587264},
publisher = {Association for Computing Machinery},
address = {New York, NY, USA},
url = {https://doi.org/10.1145/1576246.1531346},
doi = {10.1145/1576246.1531346},
booktitle = {ACM SIGGRAPH 2009 Papers},
articleno = {40},
numpages = {6},
location = {New Orleans, Louisiana},
series = {SIGGRAPH '09}
}

@article{shapiro1996asph,
  title   = {Adaptive Smoothed Particle Hydrodynamics, with Application to Cosmology: Methodology},
  author  = {Shapiro, Paul R. and Martel, Hugo and Villumsen, Jens V.},
  journal = {The Astrophysical Journal Supplement Series},
  volume  = {103},
  pages   = {269--330},
  year    = {1996},
  doi     = {10.1086/192279}
}

@article{owen1998asph2,
  title   = {Adaptive Smoothed Particle Hydrodynamics: Methodology. {II}.},
  author  = {Owen, J. Michael and Villumsen, Jens V. and Shapiro, Paul R. and Martel, Hugo},
  journal = {The Astrophysical Journal Supplement Series},
  volume  = {116},
  number  = {2},
  pages   = {155--209},
  year    = {1998},
  doi     = {10.1086/313100}
}

@article{yu2013anisotropickernel,
  title   = {Reconstructing Surfaces of Particle-Based Fluids Using Anisotropic Kernels},
  author  = {Yu, Jihun and Turk, Greg},
  journal = {ACM Transactions on Graphics},
  volume  = {32},
  number  = {1},
  year    = {2013},
  doi     = {10.1145/2421636.2421641},
  url     = {https://dl.acm.org/doi/10.1145/2421636.2421641}
}

@article{kerbl2023gaussiansplatting,
  title   = {3D Gaussian Splatting for Real-Time Radiance Field Rendering},
  author  = {Kerbl, Bernhard and Kopanas, Georgios and Leimk{\"u}hler, Thomas and Drettakis, George},
  journal = {ACM Transactions on Graphics},
  volume  = {42},
  number  = {4},
  year    = {2023},
  doi     = {10.1145/3592433},
  url     = {https://dl.acm.org/doi/10.1145/3592433}
}

@InProceedings{toshev2024neuralsph,
  title = 	 {Neural {SPH}: Improved Neural Modeling of Lagrangian Fluid Dynamics},
  author =       {Toshev, Artur and Erbesdobler, Jonas A. and Adams, Nikolaus A. and Brandstetter, Johannes},
  booktitle = 	 {Proceedings of the 41st International Conference on Machine Learning},
  pages = 	 {48428--48452},
  year = 	 {2024},
  editor = 	 {Salakhutdinov, Ruslan and Kolter, Zico and Heller, Katherine and Weller, Adrian and Oliver, Nuria and Scarlett, Jonathan and Berkenkamp, Felix},
  volume = 	 {235},
  series = 	 {Proceedings of Machine Learning Research},
  month = 	 {21--27 Jul},
  publisher =    {PMLR},
  url = 	 {https://proceedings.mlr.press/v235/toshev24a.html}
}

@inproceedings{ummenhofer2020continuousconv,
  title     = {Lagrangian Fluid Simulation with Continuous Convolutions},
  author    = {Ummenhofer, Benjamin and Prantl, Lukas and Thuerey, Nils and Koltun, Vladlen},
  booktitle = {International Conference on Learning Representations (ICLR)},
  year      = {2020},
  url       = {https://openreview.net/forum?id=B1lDoJSYDH}
}

@InProceedings{GNS,
  title = 	 {Learning to Simulate Complex Physics with Graph Networks},
  author =       {Sanchez-Gonzalez, Alvaro and Godwin, Jonathan and Pfaff, Tobias and Ying, Rex and Leskovec, Jure and Battaglia, Peter},
  booktitle = 	 {Proceedings of the 37th International Conference on Machine Learning},
  pages = 	 {8459--8468},
  year = 	 {2020},
  editor = 	 {III, Hal Daumé and Singh, Aarti},
  volume = 	 {119},
  series = 	 {Proceedings of Machine Learning Research},
  month = 	 {13--18 Jul},
  publisher =    {PMLR},
  url = 	 {https://proceedings.mlr.press/v119/sanchez-gonzalez20a.html}
}

@article{NeuralUpFlow,
author = {Roy, Bruno and Poulin, Pierre and Paquette, Eric},
title = {Neural UpFlow: A Scene Flow Learning Approach to Increase the Apparent Resolution of Particle-Based Liquids},
year = {2021},
issue_date = {September 2021},
publisher = {Association for Computing Machinery},
address = {New York, NY, USA},
volume = {4},
number = {3},
url = {https://doi.org/10.1145/3480147},
doi = {10.1145/3480147},
journal = {Proc. ACM Comput. Graph. Interact. Tech.},
month = sep,
articleno = {40},
numpages = {26}
}

@article{um2018mlflip,
  title     = {Liquid Splash Modeling with Neural Networks},
  author    = {Um, Kiwon and Hu, Xiangyu and Thuerey, Nils},
  journal   = {Computer Graphics Forum},
  volume    = {37},
  number    = {8},
  pages     = {171--182},
  year      = {2018},
  publisher = {The Eurographics Association and John Wiley \& Sons Ltd.},
  doi       = {10.1111/cgf.13522},
  url       = {https://doi.org/10.1111/cgf.13522}
}

@inproceedings{prantl2022dmcf,
  title     = {Guaranteed Conservation of Momentum for Learning Particle-based Fluid Dynamics},
  author    = {Prantl, Lukas and Ummenhofer, Benjamin and Koltun, Vladlen and Thuerey, Nils},
  booktitle = {Advances in Neural Information Processing Systems},
  volume    = {35},
  year      = {2022},
  publisher = {Neural Information Processing Systems Foundation},
  url       = {https://proceedings.neurips.cc/paper_files/paper/2022/hash/2dd7f33ffbb59b4ff987be5442a13016-Abstract-Conference.html}
}

@article{well,
  title={The well: a large-scale collection of diverse physics simulations for machine learning},
  author={Ohana, Ruben and McCabe, Michael and Meyer, Lucas and Morel, Rudy and Agocs, Fruzsina and Beneitez, Miguel and Berger, Marsha and Burkhart, Blakesly and Dalziel, Stuart and Fielding, Drummond and others},
  journal={Advances in Neural Information Processing Systems},
  volume={37},
  pages={44989--45037},
  year={2024}
}

@article{fluidgaussian,
  title={FluidGaussian: Propagating Simulation-Based Uncertainty Toward Functionally-Intelligent 3D Reconstruction},
  author={Liu, Yuqiu and Song, Jialin and Ramirez de Chanlatte, Marissa and Chowdhury, Rochishnu and Desai, Rushil Paresh and Chen, Wuyang and Martin, Daniel and Mahoney, Michael},
  journal={CVPR},
  year={2026}
}

@article{neuralmpm,
  title={Hybrid Neural-MPM for Interactive Fluid Simulations in Real-Time},
  author={Xu, Jingxuan and Huang, Hong and Zou, Chuhang and Savva, Manolis and Wei, Yunchao and Chen, Wuyang},
  journal={arXiv preprint arXiv:2505.18926},
  year={2025}
}

@article{extmps,
author = {Enomoto, Keigo and Ishida, Takato and Doi, Yuya and Uneyama, Takashi and Masubuchi, Yuichi},
title = {Extension of moving particle simulation by introducing rotational degrees of freedom for dilute fiber suspensions},
journal = {International Journal for Numerical Methods in Fluids},
volume = {96},
number = {2},
pages = {125-137},
doi = {https://doi.org/10.1002/fld.5235},
url = {https://onlinelibrary.wiley.com/doi/abs/10.1002/fld.5235},
eprint = {https://onlinelibrary.wiley.com/doi/pdf/10.1002/fld.5235},
year = {2024}
}

@article{SPH,
    author = {Gingold, R. A. and Monaghan, J. J.},
    title = {Smoothed particle hydrodynamics: theory and application to non-spherical stars},
    journal = {Monthly Notices of the Royal Astronomical Society},
    volume = {181},
    number = {3},
    pages = {375-389},
    year = {1977},
    month = {12},
    issn = {0035-8711},
    doi = {10.1093/mnras/181.3.375},
    url = {https://doi.org/10.1093/mnras/181.3.375},
    eprint = {https://academic.oup.com/mnras/article-pdf/181/3/375/3104055/mnras181-0375.pdf},
}

@inproceedings{particle_fluid_sim,
author = {M\"{u}ller, Matthias and Charypar, David and Gross, Markus},
title = {Particle-based fluid simulation for interactive applications},
year = {2003},
isbn = {1581136595},
publisher = {Eurographics Association},
address = {Goslar, DEU},
booktitle = {Proceedings of the 2003 ACM SIGGRAPH/Eurographics Symposium on Computer Animation},
pages = {154–159},
numpages = {6},
location = {San Diego, California},
series = {SCA '03}
}

@article{high_resolution_fluid_sim,
author = {Nielsen, Michael B. and Bridson, Robert},
title = {Guide shapes for high resolution naturalistic liquid simulation},
year = {2011},
issue_date = {July 2011},
publisher = {Association for Computing Machinery},
address = {New York, NY, USA},
volume = {30},
number = {4},
issn = {0730-0301},
url = {https://doi.org/10.1145/2010324.1964978},
doi = {10.1145/2010324.1964978},
journal = {ACM Trans. Graph.},
month = jul,
articleno = {83},
numpages = {8}
}

@article{CO_FLIP,
author = {Nabizadeh, Mohammad Sina and Roy-Chowdhury, Ritoban and Yin, Hang and Ramamoorthi, Ravi and Chern, Albert},
title = {Fluid Implicit Particles on Coadjoint Orbits},
year = {2024},
issue_date = {December 2024},
publisher = {Association for Computing Machinery},
address = {New York, NY, USA},
volume = {43},
number = {6},
issn = {0730-0301},
url = {https://doi.org/10.1145/3687970},
doi = {10.1145/3687970},
journal = {ACM Trans. Graph.},
month = nov,
articleno = {270},
numpages = {38}
}

@article {KBST2022,
journal = {Computer Graphics Forum},
title = {{A Survey on SPH Methods in Computer Graphics}},
author = {Koschier, Dan and Bender, Jan and Solenthaler, Barbara and Teschner, Matthias},
year = {2022},
volume ={41},
number = {2},
publisher = {The Eurographics Association and John Wiley & Sons Ltd.},
ISSN = {1467-8659},
DOI = {10.1111/cgf.14508}
}

@inproceedings{neuralfluid,
 author = {Wang, Haoxiang and Yu, Tao and Qiao, Hui and Dai, Qionghai},
 booktitle = {International Conference on Learning Representations},
 editor = {Y. Yue and A. Garg and N. Peng and F. Sha and R. Yu},
 pages = {99091--99118},
 title = {Neural Fluid Simulation on Geometric Surfaces},
 url = {https://proceedings.iclr.cc/paper_files/paper/2025/file/f58c24798220ba724fe05c0fa786227d-Paper-Conference.pdf},
 volume = {2025},
 year = {2025}
}

@article{MPS,
author = {S. Koshizuka and Y. Oka},
title = {Moving-Particle Semi-Implicit Method for Fragmentation of Incompressible Fluid},
journal = {Nuclear Science and Engineering},
volume = {123},
number = {3},
pages = {421--434},
year = {1996},
publisher = {Taylor \& Francis},
doi = {10.13182/NSE96-A24205},
URL = {https://doi.org/10.13182/NSE96-A24205},
eprint = {https://doi.org/10.13182/NSE96-A24205}
}

@article{APIC,
author = {Jiang, Chenfanfu and Schroeder, Craig and Selle, Andrew and Teran, Joseph and Stomakhin, Alexey},
title = {The affine particle-in-cell method},
year = {2015},
issue_date = {August 2015},
publisher = {Association for Computing Machinery},
address = {New York, NY, USA},
volume = {34},
number = {4},
issn = {0730-0301},
url = {https://doi.org/10.1145/2766996},
doi = {10.1145/2766996},
journal = {ACM Trans. Graph.},
month = jul,
articleno = {51},
numpages = {10}
}

@article{MLS_MPM,
author = {Hu, Yuanming and Fang, Yu and Ge, Ziheng and Qu, Ziyin and Zhu, Yixin and Pradhana, Andre and Jiang, Chenfanfu},
title = {A moving least squares material point method with displacement discontinuity and two-way rigid body coupling},
year = {2018},
issue_date = {August 2018},
publisher = {Association for Computing Machinery},
address = {New York, NY, USA},
volume = {37},
number = {4},
issn = {0730-0301},
url = {https://doi.org/10.1145/3197517.3201293},
doi = {10.1145/3197517.3201293},
journal = {ACM Trans. Graph.},
month = jul,
articleno = {150},
numpages = {14}
}

@article{MPM,
title = {Application of a particle-in-cell method to solid mechanics},
journal = {Computer Physics Communications},
volume = {87},
number = {1},
pages = {236-252},
year = {1995},
note = {Particle Simulation Methods},
issn = {0010-4655},
doi = {https://doi.org/10.1016/0010-4655(94)00170-7},
url = {https://www.sciencedirect.com/science/article/pii/0010465594001707},
author = {Deborah Sulsky and Shi-Jian Zhou and Howard L. Schreyer}
}

@article{DPD,
    author = {Groot, Robert D. and Warren, Patrick B.},
    title = {Dissipative particle dynamics: Bridging the gap between atomistic and mesoscopic simulation},
    journal = {The Journal of Chemical Physics},
    volume = {107},
    number = {11},
    pages = {4423-4435},
    year = {1997},
    month = {09},
    issn = {0021-9606},
    doi = {10.1063/1.474784},
    url = {https://doi.org/10.1063/1.474784},
    eprint = {https://pubs.aip.org/aip/jcp/article-pdf/107/11/4423/19289351/4423_1_online.pdf},
}

@article{jax-sph,
  title={JAX-SPH: A Differentiable Smoothed Particle Hydrodynamics Framework},
  author={Toshev, Artur P and Ramachandran, Harish and Erbesdobler, Jonas A and Galletti, Gianluca and Brandstetter, Johannes and Adams, Nikolaus A},
  journal={arXiv preprint arXiv:2403.04750},
  year={2024}
}

@article{ladicky2015regressionforests,
  title = {Data-driven Fluid Simulations Using Regression Forests},
  author = {Ladick{\'y}, L'ubor and Jeong, SoHyeon and Solenthaler, Barbara and Pollefeys, Marc and Gross, Markus},
  journal = {ACM Transactions on Graphics},
  volume = {34},
  number = {6},
  year = {2015},
  doi = {10.1145/2816795.2818129}
}
